%% file: paper.tex
\pdfoutput=1

\documentclass[11pt]{article}

\usepackage[preprint]{acl}

\usepackage{times}
\usepackage{latexsym}
\usepackage{enumitem}

\usepackage[T1]{fontenc}
\usepackage[utf8]{inputenc}
\usepackage{microtype}
\usepackage{inconsolata}
\usepackage{graphicx}
\usepackage{amsmath,amssymb,amsthm}
\usepackage{algorithm}
\usepackage{algpseudocode}

\newtheorem{definition}{Definition}

\usepackage{xcolor}
\newcommand{\meq}[1]{{\texttt{#1}}}
\usepackage{todonotes}
\input{sections/macros}

\title{LLM-based Cognitive Models of Students with Misconceptions}

\author{
  Shashank Sonkar\\
  Rice University \\
  \texttt{ss164@rice.edu} \\
  \And
  Xinghe Chen\\
  Rice University \\
  \texttt{xc42@rice.edu} \\
  \And
  Naiming Liu\\
  Rice University \\
  \texttt{nl35@rice.edu} \\
  \AND 
  Richard G. Baraniuk\\
  Rice University \\
  \texttt{richb@rice.edu} \\
  \And
  Mrinmaya Sachan \\
  ETH Zurich \\
  \texttt{mrinmaya.sachan@inf.ethz.ch} \\
}

\begin{document}
\maketitle
\begin{abstract}
\input{sections/abstract}
\end{abstract}

\input{tables/pt_w_arch}
\section{Introduction}
\input{sections/intro}

\section{Related Work}
\input{sections/related_work}

\input{figures/algos}
\section{MalAlgoPy: Framework for Algebraic Problem-Solving with Misconceptions}
\input{sections/library}

\section{Cognitive Student Models}
\input{sections/method}

\input{figures/split_AND_strong_csa}
\section{Instruction-tuning LLMs as CSMs}
\input{sections/experiments}

\section{Conclusion}
\input{sections/conclusion}

\section{Limitations}
\input{sections/limitations}

\bibliography{custom}

\newpage
\onecolumn
\appendix
\section{Appendix}
\label{sec:appendix}
\input{sections/appendix}
\end{document}

%% file: sections/macros.tex
\definecolor{TodoColor}{rgb}{1,0.7,0.6}
\definecolor{TodoColor2}{HTML}{AACCAA}

\makeatletter\def\Hy@Warning#1{}\makeatother
\let\svthefootnote\thefootnote
\newcommand\blankfootnote[1]{%
  \let\thefootnote\relax\footnotetext{#1}%
  \let\thefootnote\svthefootnote%
}

%% file: sections/abstract.tex
Accurately modeling student cognition is crucial for developing effective AI-driven educational technologies. A key challenge is creating realistic student models that satisfy two essential properties: (1) accurately replicating specific misconceptions, and (2) correctly solving problems where these misconceptions are not applicable. This dual requirement reflects the complex nature of student understanding, where misconceptions coexist with correct knowledge.
This paper investigates whether Large Language Models (LLMs) can be instruction-tuned to meet this dual requirement and effectively simulate student thinking in algebra. We introduce MalAlgoPy, a novel Python library that generates datasets reflecting authentic student solution patterns through a graph-based representation of algebraic problem-solving. Utilizing MalAlgoPy, we define and examine Cognitive Student Models (CSMs) – LLMs instruction tuned to faithfully emulate realistic student behavior.
Our findings reveal that LLMs trained on misconception examples can efficiently learn to replicate errors. However, the training diminishes the model's ability to solve problems correctly, particularly for problem types where the misconceptions are not applicable, thus failing to satisfy second property of CSMs. We demonstrate that by carefully calibrating the ratio of correct to misconception examples in the training data – sometimes as low as 0.25 – it is possible to develop CSMs that satisfy both properties. Our insights enhance our understanding of AI-based student models and pave the way for effective adaptive learning systems.

%% file: tables/pt_w_arch.tex
\begin{figure*}[t]
\begin{minipage}{0.3\textwidth}
\centering
\resizebox{0.9\linewidth}{!}{
\begin{tabular}{|l|l|}
\hline
\textbf{Type} & \textbf{Expression} \\
\hline
T1 & $Ax = B$ \\
T2 & $Ax = B + C$ \\
T3 & $Ax = B * C$ \\
T4 & $Ax + Bx = C$ \\
T5 & $Ax + B = C$ \\
T6 & $A + Bx = C$ \\
T7 & $Ax = Bx + C$ \\
T8 & $Ax = B(C*D)$ \\
T9 & $Ax = B(Cx + D)$ \\
T10 & $Ax = B + C * D$ \\
T11 & $A + Bx + Cx = D$ \\
T12 & $Ax = B + C(Dx + E)$ \\
T14 & $Ax + B = Cx + D$ \\
T15 & $Ax + Bx = C + D$ \\
T16 & $Ax = Bx + C + D$ \\
\hline
\end{tabular}
}
\label{tab:problem_types}
\end{minipage}%
\begin{minipage}{0.7\textwidth}
\centering
\includegraphics[width=\linewidth]{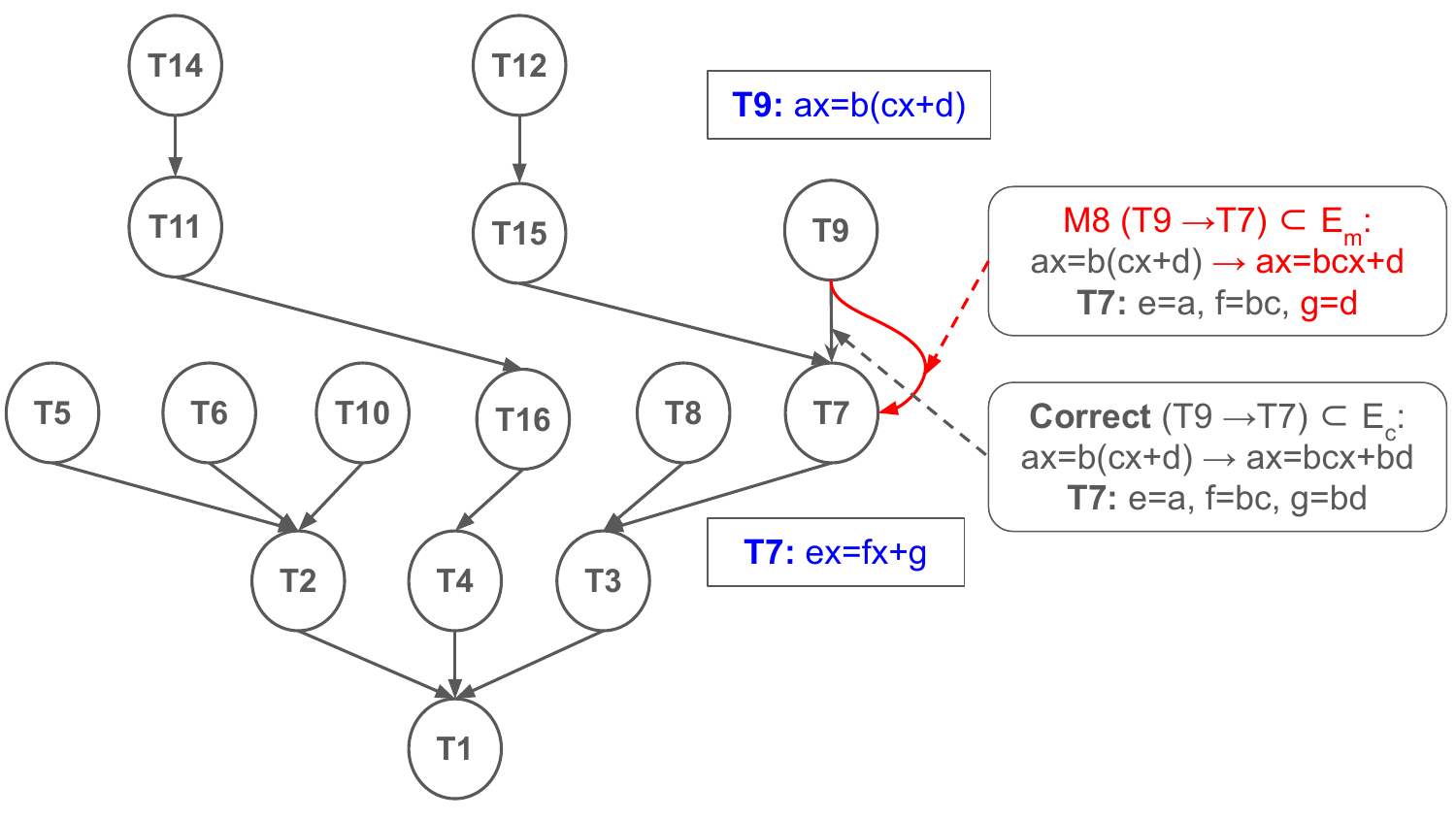}
\end{minipage}
\vspace{-3mm}
\caption{
\textbf{Left:} Problem types and corresponding algebraic expressions. 
\textbf{Right:} MalAlgoPy's graph-based model of algebraic problem-solving.
This directed acyclic graph $G = (V, E_C \cup E_M)$ models cognitively faithful step-by-step simplifications as performed by students solving linear equations. 
$V = \{T1, \ldots, T16\}$ denotes problem types, with T1 representing the base case $Ax = B$. $E_C \subset V \times V$ represents correct transformations (grey arrows), while $E_M \subset V \times V$ represents misconception transformation (red arrows). The graph topology encodes the solution space, with all paths converging to T1. All possible correct reductions are shown as grey arrows, while only one example misconception is depicted with red arrows for clarity. For instance, T9 \meq{(ax=b(cx+d))} can correctly reduce to T7 \meq{(ex=fx+g)} where $e=a$, $f=bc$, $g=bd$ (grey arrow), or incorrectly reduce via misconception M8  (distributive law applied only to the first term) where $g=d$ (red arrow). This structure enables generation of both correct solutions and ``malgorithms'' that reflect common student solutions.
}
\vspace{-5mm}
\label{fig:tree}
\end{figure*}

%% file: sections/intro.tex
The intersection of artificial intelligence (AI) and education is ushering in a new era of possibilities for how we understand, support, and enhance student learning. At the heart of this transformation is the concept of student modeling, which refers to the creation of computational representations of students' knowledge, skills, and misconceptions \cite{corbett1994knowledge}. Accurate student models serve multiple critical functions. They enable adaptive testing, efficiently assessing student knowledge by tailoring test items to the student's demonstrated abilities \cite{baker_test}. These models are integral to Intelligent Tutoring Systems, enabling personalized feedback and adaptive instruction \cite{pardos2013adapting}. Additionally, student models aid in training teachers using simulated learning environments \cite{piech2023gpteach}.

A significant challenge in student modeling is the accurate representation of misconceptions. Misconceptions are not random errors; they are systematic, persistent, predictable patterns reflecting underlying conceptual misunderstandings \cite{malrules}. 
For example in Figure~\ref{fig:tree}, when solving $ax=b(cx+d)$, a common misconception leads students to incorrectly distribute $b$ to only the first term, resulting in $ax=bcx+d$. While general feedback might state, ``Your solution is incorrect. Please review the steps,'' targeted feedback addresses the specific misconception: ``In $ax=b(cx+d)$, $b$ multiplies both terms inside the parentheses: $ax = b(cx) + b(d) = bcx + bd$.'' Such targeted feedback helps students correct their fundamental conceptual misunderstandings.

Recent advancements in Large Language Models (LLMs) offer promising opportunities to enhance student modeling in education. While LLMs excel in natural language tasks, their ability to accurately model student cognition, particularly in replicating misconceptions, remains largely unexplored. This gap presents a significant opportunity: if LLMs can be trained to accurately simulate student thinking as Cognitive Student Models (CSMs), they could revolutionize education through more personalized and effective learning experiences.

To systematically study LLMs' capacity to model CSMs, we developed MalAlgoPy, a novel Python library that represents algebraic equations and their solution processes as a directed acyclic graph (Figure~\ref{fig:tree}).
MalAlgoPy models 16 distinct problem types as nodes in the graph, with edges representing both correct transformations (grey arrows) and 20 common misconceptions (red arrows - only one shown). 
These misconceptions were diagnosed as common algebraic errors in secondary school students, providing a realistic basis for modeling student thinking \cite{sleeman,sleeman2,malrules}.
MalAlgoPy handles a broad range of linear equations with one variable, including complex forms with parentheses and multiple terms. This structure enables the generation of diverse, cognitively-grounded datasets capturing both correct solution paths and ``malgorithms''  - solution paths incorporating common student errors. Using MalAlgoPy, we can conduct a comprehensive study on training LLMs to simulate student cognitive processes in algebra.

First, we formally define a Cognitive Student Model (CSM) as follows: A CSM model $A$ satisfies two properties: (1). Given a misconception $M$ and problem types where it's applicable, $A$ accurately applies the misconception; (2). For problem types where $M$ doesn't apply, $A$ solves problems correctly. Both these properties are important for CSMs to faithfully represent the nuanced nature of student understanding, where misconceptions coexist with correct knowledge across different problem contexts. The importance of these dual requirements becomes clear when considering CSMs' application in adaptive testing. By accurately replicating both misconception application and correct problem-solving, CSMs enable precise mapping of a student's knowledge boundaries. This nuanced representation enables targeted question selection, leading to a faster, more accurate assessment of student proficiency.

Our experimentation process begins by instruction-tuning Llama models \citep{llama3} to create CSMs. We train models on misconception examples, uncovering intriguing dynamics in how LLMs acquire misconceptions \cite{power2022grokking}. While these models learn to apply misconceptions accurately with just 400-800 examples—fewer than the 2000 typically required for mastering correct solutions—we observe that they only satisfy the first property of CSMs. These models exhibit diminished ability to solve problems correctly for types where learned misconceptions are not applicable, thus failing to meet the second property of CSMs.

\input{tables/misc_2_pt_map}

To address this limitation and satisfy both properties of CSMs, we investigate strategies to balance misconception acquisition with preserving correct problem-solving abilities. By carefully calibrating the training data to include both misconception and correct examples—with ratios of correct to misconception examples sometimes as low as 0.25—we achieve models that meet both requirements of CSMs. This novel training paradigm successfully creates CSMs that faithfully represent student understanding.

Our contributions in this work are foundational and pave the way for future research: the development of MalAlgoPy, the introduction of a new task of Cognitive Student Modeling, and a focused investigation into how LLMs can be instruction-tuned to simulate these models.

%% file: tables/misc_2_pt_map.tex
\begin{table*}[th!]
\centering
\resizebox{\textwidth}{!}{
\begin{tabular}{|l|l|l|l|}
\hline
\textbf{Misconception} & \textbf{Expression} & \textbf{Applicable Types} & \textbf{Description} \\
\hline
M1 & $A(part) \to A + (part)$ & T8, T9, T10, T12 & Treating distribution as addition \\
M2\_S3 & $A(Bx \pm C) \to ABx \pm C$ & T9, T12 & Ignoring distribution \\
M3 & $A \pm B(part) \to (A \pm B)(part)$ & T10, T12 & Misapplying parentheses \\
M4 & $A(B*C) \to A*B*A*C$ & T8 & Incorrectly distributing multiplication \\
M5 & $A(Bx \pm C) \to A(A*Bx \pm A*C)$ & T9, T12 & Over-distribution \\
M6 & $-A(Bx - C) \to -A*Bx - A*C$ & T9, T12 & Incorrect sign distribution \\
M8 & $A(Bx \pm C) \to Bx \pm A*C$ & T9, T12 & Incorrect distribution on x term \\
M11 & $Ax \pm B = Cx \pm D \to Ax + Cx = B + D$ & T14 & Incorrectly combining terms \\
M12/S15 & $Ax \pm B = (A \pm B)x$ & T5, T6, T7, T9, T12 & Incorrectly factoring x \\
M13 & $Ax \pm B = (A \pm B)$ & T5, T6, T7, T9, T12 & Incorrectly factoring x \\
M14 & $part1 + part2 \to part1 - part2$ & T2, T4 & Incorrectly swapping addition and subtraction \\
M15 & $part1 - part2 \to part1 + part2$ & T2, T4 & Incorrectly swapping addition and subtraction \\
M16 & $part1 * part2 \to part1 + part2$ & T3, T10 & Treating multiplication as addition \\
M17 & $A + B \to B - A$ & T2, T4 & Incorrectly swapping order of addition and subtraction \\
M18 & $A - B \to B - A$ & T2, T4 & Incorrectly swapping order of addition and subtraction \\
M19 & $Ax = B \to x = A + B$ & All Types & Treat division as addition \\
M20\_S20 & $Ax = B \to x = B$ & All Types & Divide only on one side \\
M21 & $Ax = B \to x = A - B$ & All Types & Treat division as subtraction \\
M22\_S1 & $Ax = B \to x = A/B$ & All Types & Incorrect numerator and denominator \\
\hline
\end{tabular}
}
\vspace{-2mm}
\caption{Overview of algebraic misconceptions modeled in MalAlgoPy. Each row presents a specific misconception, its mathematical expression, the problem types to which it applies, and a brief description of the error.}
\label{tab:misc}
\vspace{-4mm}
\end{table*}

%% file: sections/related_work.tex
\label{sec:related}
Cognitive modeling in mathematics education has significantly evolved since its inception. The groundwork was laid by seminal works such as problem space theory \citep{newell1972human} and ACT-R theory \citep{anderson1993rules}, which emphasized understanding step-by-step cognitive processes in problem-solving. Early computational approaches like BUGGY \citep{brown1978diagnostic} employed rule-based systems for modeling arithmetic errors. This paved the way for sophisticated models like the MalRules approach \citep{malrules}, which focused on modeling algebraic misconceptions using both correct rules and ``malrules'' to capture common student errors. Representation of student misconceptions has also been critical in cognitive modeling in mathematics education. \citealp{matz1982towards} provided an analysis of systematic errors in algebra, categorizing them into extrapolation, generalization, and repairing errors. The application of cognitive models in intelligent tutoring systems like the Cognitive Tutor \citep{ritter2007cognitive} demonstrates the effectiveness of cognitive modeling in providing personalized instruction.

%% file: figures/algos.tex
\begin{figure*}[th!]
\centering
\begin{minipage}[t]{0.32\textwidth}
\begin{algorithm}[H]
\small
\caption{Reduction Process}
\label{fig:algo1}
\begin{algorithmic}[1]
\Function{Reduce}{$e, T_i$}
    \If{$T_i = T_1$}
        \Return $e$
    \EndIf
    \State $e' \gets R_i(e)$
    \State $T_j \gets$ type of $e'$
    \If{$(T_i, T_k) \in E_M$ for some $T_k$}
        \State Choose between $(T_i, T_j)$ and $(T_i, T_k)$
        \State Update $e'$ and $T_j$ if misconception is chosen
    \EndIf
    \State \Return \Call{Reduce}{$e', T_j$}
\EndFunction
\end{algorithmic}
\end{algorithm}
\end{minipage}
\hfill
\begin{minipage}[t]{0.59\textwidth}
\begin{algorithm}[H]
\small
\caption{Reduction with Misconceptions}
\label{fig:algo2}
\begin{algorithmic}[1]
\Function{ReduceWithMisconceptions}{$e, T_i, \mathcal{M}$}
    \If{$T_i = T_1$}
        \Return $e$
    \EndIf
    \State $e' \gets R_i(e)$
    \State $T_j \gets$ type of $e'$
    \For{$M \in \mathcal{M}$}
        \If{$M(T_i, T_j) \in E_M$}
            \State $(T_i, T_k) \gets M(T_i, T_j)$
            \State $e'' \gets$ apply $M$ to $e'$
            \State \Return \Call{ReduceWithMisconceptions}{$e'', T_k, \mathcal{M}$}
        \EndIf
    \EndFor
    \State \Return \Call{ReduceWithMisconceptions}{$e', T_j, \mathcal{M}$}
\EndFunction
\end{algorithmic}
\end{algorithm}
\end{minipage}
\vspace{-2mm}
\caption{Algorithms for algebraic equation reduction in MalAlgoPy: (left) Basic reduction process that transforms equations to simpler forms; (right) Extended reduction process that systematically applies a set of predefined misconceptions, generating both correct and incorrect solution paths. $\mathcal{M}$ is the set of all applicable misconceptions.}
\label{fig:algorithms}
\vspace{-5mm}
\end{figure*}

%% file: sections/library.tex
\label{sec:malalgopy}
MalAlgoPy's architecture is built on three core design principles that work together to model algebraic problem-solving and student misconceptions. These principles ensure extensibility, modularity, and faithful representation of student cognitive processes in mathematical reasoning.

\subsection{Problem Type Hierarchy}
\label{sec:core_intro}

MalAlgoPy is designed to handle linear equations with one variable, including more complex forms with parentheses and multiple terms.
At its core is a type-based problem representation implemented as a directed acyclic graph $G = (V, E)$.
The set $V = \{T_1, T_2, \ldots, T_{16}\}$ represents a hierarchy of problem types, where each $T_i \in V$ corresponds to a specific equation structure.
For instance, $T_9$ represents equations of the form \meq{ax = b(cx + d)}, while $T_7$ represents equations of the form \meq{ex = fx + g}. This design has three key advantages (i) it enhances code reusability across different equation types; (ii) it facilitates the integration of new equation types into the existing framework, and (iii) it closely mirrors the student cognitive process of algebraic simplification.

The set $E = E_C \cup E_M$ represents transitions between types, where $E_C$ denotes correct reductions and $E_M$ represents misconception-based transitions. Correct reductions, represented by edges in $E_C$, are valid transformations that simplify equations. For example, an edge $(T_9, T_7) \in E_C$ represents the correct reduction of $ax = b(cx + d)$ to $ax = bcx + bd$, where $e = a$, $f = bc$, and $g = bd$. 

Misconception-based transformations, represented by a set of edges in $E_M$, model common student errors in equation transformations. As shown in Figure~\ref{fig:tree}, when solving $ax = b(cx + d)$, a common misconception (M8) leads students to incorrectly distribute $b$ to only the first term. This erroneous transformation is represented by an edge $(T_9, T_7) \in E_M$. A misconception can be applied to multiple problem types as shown in Table~\ref{tab:misc}. For e.g. M8 is applicable to T12: \meq{(ax=b+c(dx+e))} as well as T9, thus M8 $=\{(T_9, T_7), (T_{12}, T_{15}) \}$.

\subsection{Single-Step Reduction Process}

MalAlgoPy enforces a strict single-step reduction policy, closely mirroring the student cognitive process of step-by-step problem-solving. Each type class implements a \texttt{reduce()} method that applies a single algebraic transformation, transitioning the equation to a less complex type. 
This granular reduction approach enables:
(i) detailed tracking and explanation of each solution step;
(ii) well-defined insertion points for modeling misconceptions; and 
(iii) precise control over the solution process allowing generation of comprehensive solution trees, including both correct and erroneous paths. 

The reduction process leverages the type hierarchy to transform complex equations into progressively simpler forms. For instance, consider the reduction path of a T9 equation ($Ax = B(Cx + D)$):
T9 \meq{(ax = b(cx + d))} $\rightarrow$ T7 \meq{(ax = bx + c)} $\rightarrow$ T2 \meq{(ax = b + c)} $\rightarrow$ T1 \meq{(ax = b)}. Each transition represents a specific algebraic operation, such as distribution or combining like terms. This process continues recursively until reaching the base case of T1.
The reduction process is formalized as a traversal of $G$, starting from the node representing the input equation type and ending at $T_1$. Algorithms~\ref{fig:algo1} and \ref{fig:algo2} present the pseudocode for the core reduction process and the integration of misconceptions, respectively.

\subsection{Decoupled Misconception Modeling}

The third principle of MalAlgoPy is decoupled misconception modeling, where misconceptions are implemented as alternative edges in the graph, representing direct transitions between problem types. This approach allows for flexible and extensible modeling of student errors within the graph-based framework. It enables cross-type applicability, where a single misconception can be applied across multiple equation types, and combinatorial error generation, where the graph structure allows for the natural combination of multiple misconceptions along a solution path. Additionally, this design facilitates easy extensibility, as new misconceptions can be added by introducing new edges in $E_M$ without modifying existing problem-type implementations, supporting ongoing research and refinement of the error model.

These three principles enable MalAlgoPy to generate a comprehensive representation of student problem-solving possibilities. The system can perform an exhaustive search, generating all possible combinations of correct steps and misconceptions for a given problem. This produces a complete solution space that captures the diverse ways students might approach an algebraic problem.

%% file: sections/method.tex
\label{sec:csa}

Our study aims to instruct-tune LLMs as CSMs which are capable of accurately simulating student problem-solving behavior in algebraic equations, including both correct solutions and common misconceptions. We will define the problem space, introduce evaluation metrics, and then define CSMs before outlining our experimental procedure.

\subsection{Problem Space}

Let $A$ be a cognitive student model, $P = \{T_1, T_2, ..., T_n\}$ be the set of problem types, where each $T_i$ represents a specific type of algebraic equation as defined in MalAlgoPy. Let $M = \{M_1, M_2, ..., M_m\}$ be the set of misconceptions, where each $M_j$ represents a specific algebraic misconception. We define a function $\alpha: M \rightarrow \mathcal{P}(P)$ that maps each misconception to the set of problem types to which it is applicable $\alpha(M) = \{T_i \in P: M \text{ is applicable to } T_i\}$.

\subsection{Evaluation Metrics}

We define the following metrics:

\textbf{Misconception Accuracy (MA)}: Assesses a model's ability to accurately replicate a specific misconception for applicable problem types.
\begin{equation*}
\text{MA}(A, M) = \frac{1}{|\alpha(M)|} \sum_{T_i \in \alpha(M)} \text{MA}(A, M, T_i),
\end{equation*}
where $\text{MA}(A, M, T_i)$ is the accuracy of model $A$ in applying misconception $M$ to problem type $T_i$.

\textbf{Correct Accuracy (Applicable) (CA$_A$)}:
Assesses a model's ability to solve problems correctly for problem types where a given misconception is applicable.
\begin{equation*}
\text{CA}_A(A, M) = \frac{1}{|\alpha(M)|} \sum_{T_i \in \alpha(M)} \text{CA}(A, T_i).
\end{equation*}

\textbf{Correct Accuracy (Non-Applicable) (CA$_{NA}$)}:
Assesses a model's performance on problem types not applicable to a given misconception.
\begin{equation*}
\text{CA}_{NA}(A, M) = \frac{1}{|P \setminus \alpha(M)|} \sum_{T_i \notin \alpha(M)} \text{CA}(A, T_i).
\end{equation*}

\textbf{Overall Correct Accuracy (OCA)}:
Assesses a model's overall ability to solve problems correctly across all problem types.
\begin{equation*}
\text{OCA}(A) = \frac{1}{|P|} \sum_{T_i \in P} \text{CA}(A, T_i).
\end{equation*}

\begin{definition}[Cognitive Student Model]
A Cognitive Student Model $A$ accurately replicates misconceptions for applicable problem types while maintaining the ability to correctly solve non-applicable problem types. Thus $A$ satisfies two properties:

\noindent\textbf{Property 1:} Given a misconception $M$ and problem types where it's applicable, $A$ accurately applies the misconception:
\begin{equation*}
\text{MA}(A, M) \geq \theta_m,
\end{equation*}
\noindent\textbf{Property 2:} For problem types where $M$ doesn't apply, $A$ solves problems correctly:
\begin{equation*}
\text{CA}_{NA}(A, M) \geq \theta_c
\end{equation*}
where $\theta_m$ and $\theta_c$ are sufficiently high thresholds for misconception accuracy and correct solving accuracy, respectively (both set to $90.0$ in experiments).
\end{definition}

\input{figures/misc_acc}

\subsection{Experimental Procedure}
Our experiment focuses on training CSMs, examining how they learn to replicate misconceptions while maintaining correct problem-solving abilities. We systematically vary the amount of training data for misconceptions and correct examples, evaluating the models' performance using our defined metrics. This process allows us to uncover learning dynamics and identify optimal training regimes for cognitively faithful student models. The procedure consists of the following steps:

\begin{enumerate}[itemsep=-2pt, parsep=0pt, left=0pt]
\item Train a base model Llama (\texttt{Llama-3.1-8B-Instruct}) on 2000 correct examples for each problem type such that $\text{OCA}(A) \geq 90.0$. This model forms the foundation for all our CSM models.

\item For each misconception $M_j$, train models $A(M_j, n_m)$ on varying amounts of misconception examples $n_m \in {100, 200, 400, 800, 1600, 3200}$. Evaluate these models to determine the relationship between $n_m$ and $\text{MA}(A, M_j)$, focusing on achieving $\text{MA}(A, M_j) > 90.0$.

\item Assess the impact of misconception training on problem-solving abilities by evaluating $\text{OCA}_(A)$, $\text{CA}_A(A, M_j)$, and $\text{CA}_{NA}(A, M_j)$ for these models.

\item To address the limitation of models trained solely on misconception examples (which satisfy property 1 but fail property 2 of CSMs), train models $A(M_j, n_m, n_c)$ by introducing correct examples $n_c$ alongside misconception examples $n_m$. Vary the ratio $n_c/n_m \in {0.25, 0.5, 1.0}$ to find the optimal balance for satisfying both CSM properties.

\item Analyze the resulting data to determine:
   \begin{enumerate}[itemsep=-1pt, parsep=0pt, left=0pt]
   \item The relationship between $n_m$ and $\text{MA}(A, M_j)$.
   \item The trade-off between $\text{MA}(A, M_j)$ and $\text{CA}_A(A, M_j)$.
   \item The optimal ratio $n_c/n_m$ for achieving high $\text{MA}(A, M_j)$ while maintaining $\text{CA}_{NA}(A, M_j) \geq \theta_c$ for CSMs.
   \end{enumerate}
\end{enumerate}

For the choice of hyperparameters, we used a learning rate of $2 \times 10^{-5}$ with AdamW \citep{loshchilov2019decoupled}, batch size of $16$, cosine scheduler, weight decay of $0.05$, warmup ratio of $0.1$, and total of $3$ epochs. For each problem type, we evaluate accuracy on $500$ test samples. This experimental setup provides a rigorous framework for uncovering the scaling laws governing the development of cognitively faithful student models.

%% file: figures/misc_acc.tex
\begin{figure*}[t!]
\centering
\includegraphics[width=0.9\linewidth]{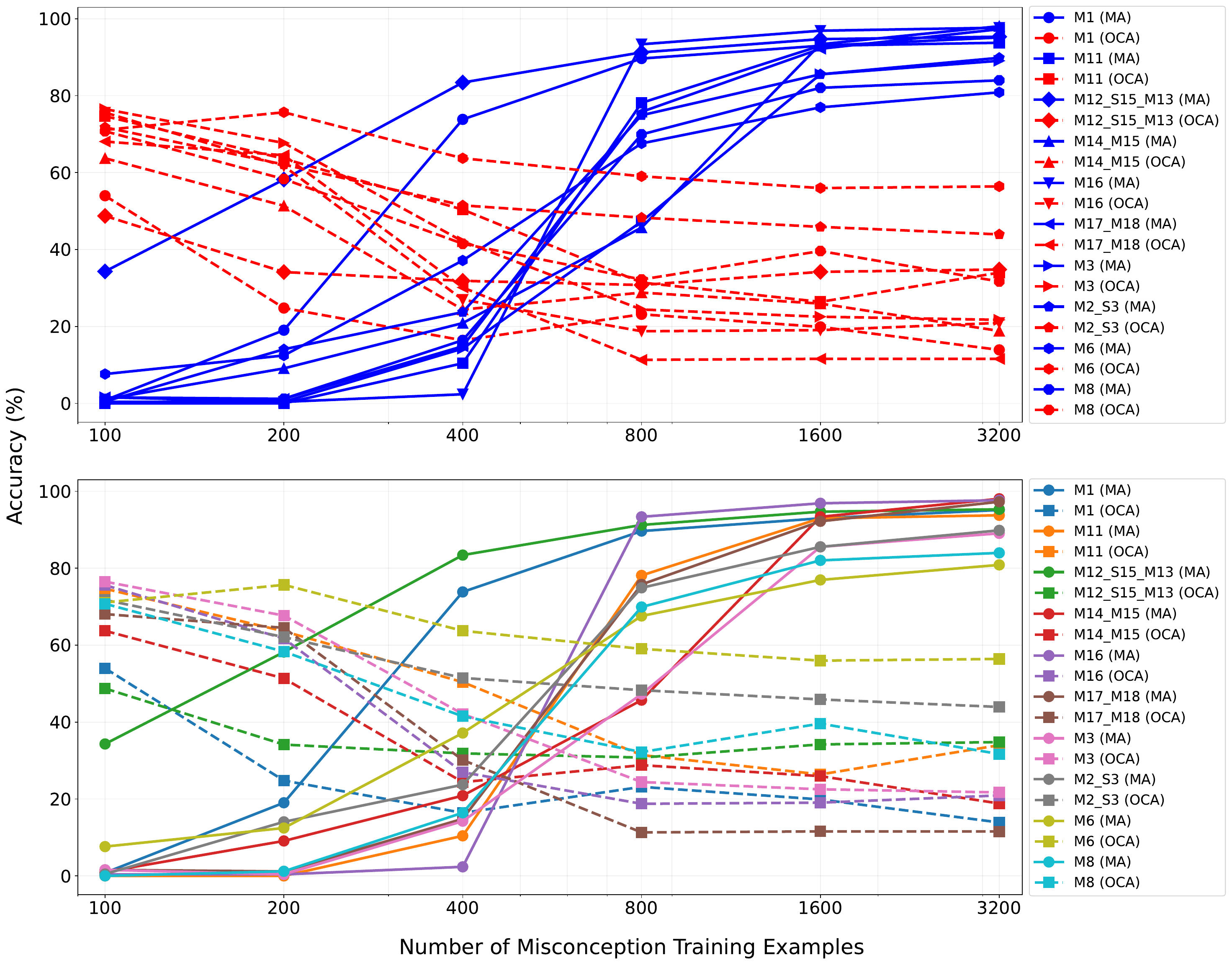} 
\vspace{-2mm}
\caption{Learning dynamics of misconceptions in Cognitive Student Models (CSMs). Each point represents a model trained on a specific number of misconception examples. As misconception training examples increase, Misconception Accuracy (MA, solid lines) rises, demonstrating an improved ability to replicate specific misconceptions for applicable problem types. Simultaneously, Overall Correct Accuracy (OCA, dashed lines) decreases, indicating a decline in the model's ability to solve problems correctly across all types. \textbf{Top:} Aggregate view of all misconceptions, with MA in blue and OCA in red. \textbf{Bottom}: Individual misconception patterns, each in a unique color. This inverse relationship between MA and OCA highlights the challenge of simultaneously satisfying both properties of CSMs: accurately replicating misconceptions while maintaining correct problem-solving abilities for non-applicable types. This inverse relationship between MA and OCA illustrates the challenge of developing CSMs.
Note that for both plots, the initial point (misconception training examples = 0) represents the base fine-tuned Llama model with OCA $\geq 90.0\%$ and MA $<5.0\%$, trained on 2000 correct examples per problem type.}
\label{fig:misconception_vs_correct_accuracy_4plots}
\vspace{-5mm}
\end{figure*}

%% file: figures/split_AND_strong_csa.tex
\begin{figure*}[t!]
\centering
\begin{minipage}[t]{0.48\textwidth}
    \centering
    \includegraphics[width=\linewidth]{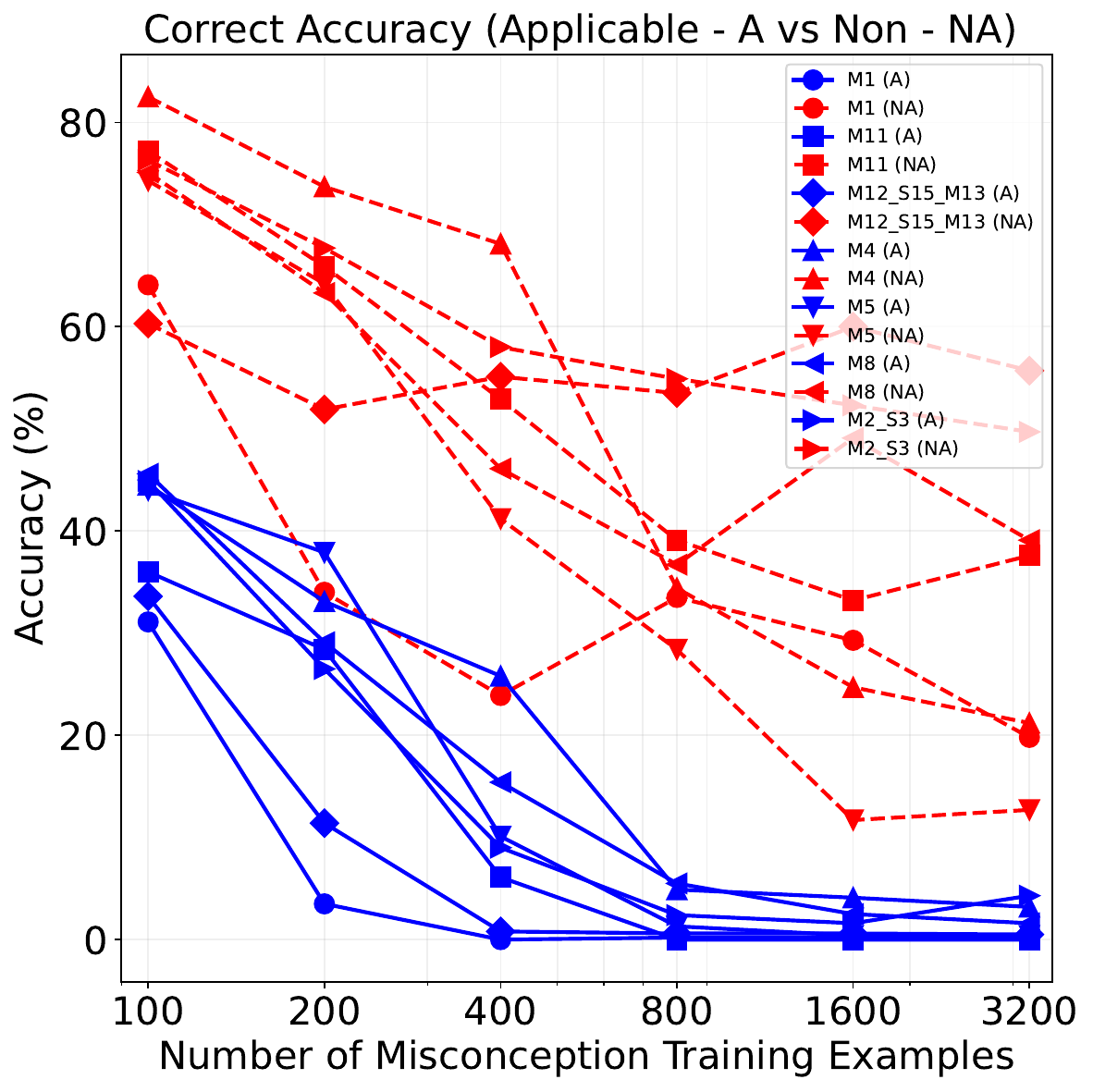}
    \label{fig:avg_accuracy_A_vs_NA_one_plot}
\end{minipage}%
\hfill
\begin{minipage}[t]{0.48\textwidth}
    \centering
    \includegraphics[width=\linewidth]{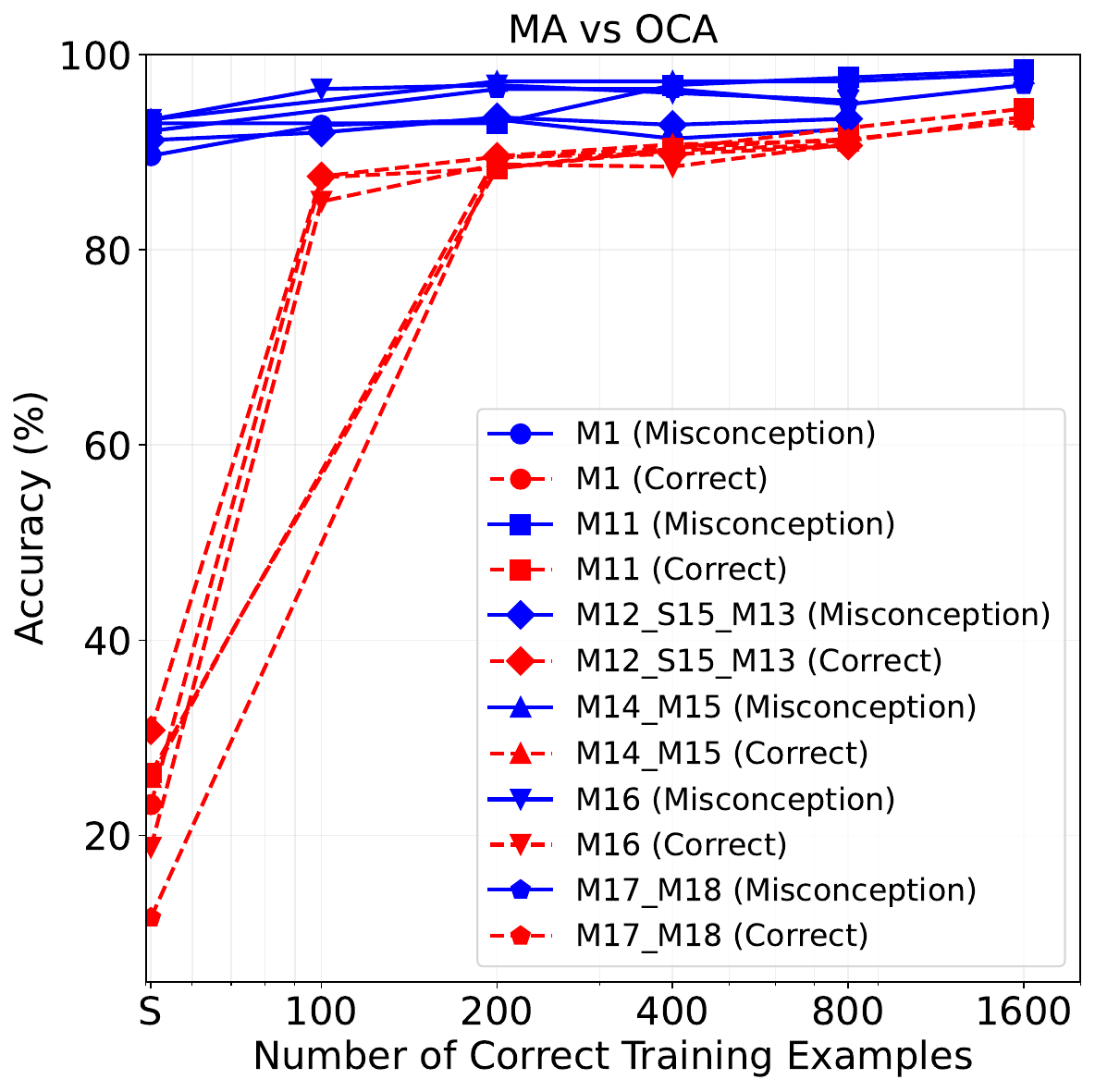}
    \label{fig:correct_solving_accuracy_one_plot}
\end{minipage}
\vspace{-7mm}
\caption{
\textbf{Left:} Impact of misconception training on correct problem-solving accuracy for CSMs. Solid lines represent accuracy for problem types where the misconception is applicable (A), while dashed lines show accuracy for non-applicable (NA) types. As the number of misconception training examples increases, accuracy decreases for both A and NA problems, with A types generally more affected. This demonstrates the challenge in training CSMs that can accurately apply misconceptions while maintaining correct problem-solving abilities for non-applicable problem types.
\textbf{Right:} Recovery of correct problem-solving abilities with introduction of correct examples. Blue lines represent Misconception Accuracy (MA), while red lines show Overall Correct Accuracy (OCA). `S' indicates the starting point with only misconception training. The graph demonstrates how increasing the number of correct training examples alongside misconception examples enables the model to balance both MA and OCA. The rapid improvement in OCA, even with a relatively small number of correct examples, suggests an efficient method for training CSMs capable of replicating misconceptions while preserving general problem-solving skills.}
\label{fig:weak_csa_a_vs_na_strong_csa}
\vspace{-4mm}
\end{figure*}

%% file: sections/experiments.tex
\label{sec:exps}

In this section, we present the results of our experiments on training Cognitive Student Models (CSMs). We examine the learning dynamics of these models, focusing on how they acquire misconceptions (property 1 of CSMs) and the consequent impact on their ability to solve problems correctly for non-applicable types (property 2 of CSMs). We then explore LLM training strategies to satisfy both properties simultaneously through the strategic introduction of correct examples.

\subsection{Misconception Acquisition in CSMs}

Our investigation began by fine-tuning the \texttt{Llama-3.1-8B-Instruct} model on 2000 correct examples for each of the 15 problem types defined in MalAlgoPy. Our objective was to achieve an Overall Correct Accuracy (OCA) of at least 90\%. This fine-tuned model forms the foundation for all CSM models, ensuring a robust initial capacity for correct problem-solving across all problem types before the introduction of misconceptions.

We then proceeded to train and evaluate CSMs $A(M_j, n_m)$ for each misconception $M_j$, systematically varying the misconception training size $n_m$. Specifically, we explored $n_m \in \{100, 200, 400, 800, 1600, 3200\}$, allowing us to observe the full spectrum of learning dynamics from initial acquisition to potential saturation effects. Our investigation revealed intriguing patterns in how these models learn to replicate algebraic misconceptions and the consequent effects on their problem-solving abilities. Figure ~\ref{fig:misconception_vs_correct_accuracy_4plots} provides a comprehensive view of these dynamics across all misconception types.

\subsubsection{Misconception Acquisition Dynamics}

Our analysis of Misconception Accuracy (MA) across different training sizes revealed intriguing patterns in how models acquire misconceptions relative to learning correct problem-solving strategies. While the base model required 2000 correct examples to achieve 90\% accuracy in correct problem-solving, the number of examples needed to learn misconceptions varied dramatically. For many misconceptions, we observed a characteristic S-shaped learning curve. Models often achieved high MA (>90\%) with an order of magnitude fewer examples than required for correct problem-solving. For instance, misconceptions like M1 (distributing multiplication over parentheses incorrectly) and M11 (incorrectly combining terms) reached over 90\% MA with 800 examples, less than half the number needed for correct problem-solving.

\subsubsection{Impact on Problem-Solving Ability}

Our analysis revealed a striking impact of misconception training on problem-solving abilities,  highlighting the challenge of simultaneously satisfying both properties of CSMs. Figure ~\ref{fig:misconception_vs_correct_accuracy_4plots} illustrates a consistent pattern across most misconceptions: as the number of training examples increases, Misconception Accuracy (MA) rises while Overall Correct Accuracy (OCA) falls dramatically. For example, with M1, the OCA dropped from 64.1\% with 100 training examples to a mere 19.8\% with 3200 examples. This inverse relationship between MA and OCA underscores a fundamental challenge in cognitive modeling: as models become more adept at replicating specific errors (property 1), they lose the ability to solve problems correctly (property 2).

To better understand this phenomenon, we examined the Correct Accuracy for Applicable ($CA_A$) and Non-Applicable ($CA_{NA}$) problem types, as shown in Figure ~\ref{fig:avg_accuracy_A_vs_NA_4plots}. We observed that as MA increased, both $CA_A$ and $CA_{NA}$ decreased, with $CA_A$ typically showing a more pronounced decline. For instance, with misconception M1, we observed a significant decline in both $CA_A$ and $CA_{NA}$. As the number of training examples increased from 100 to 3200, the $CA_A$ dropped dramatically from 31.1\% to 0.2\%. We also observed a substantial decline in $CA_{NA}$, though less severe, decreasing from 64.1\% to 19.8\% over the same range of training examples.
This pattern was consistent across various misconceptions.

\subsection{Satisfying both properties in CSMs}

The findings from our initial experiments reveal a fundamental challenge in cognitive modeling: Is it possible to create a model that satisfies both properties of CSMs simultaneously? To address this challenge, we trained and evaluated CSMs $A(M_j, n_m, n_c)$ for each misconception $M_j$. Here, $n_m$ was fixed at the number of misconception examples needed to achieve greater than 90\% misconception accuracy in the initial experiments. We then systematically varied the number of correct training examples $n_c$, exploring ratios of $n_c/n_m = \{ 0.25, 0.5,$ and $1.0 \}$. This approach allowed us to investigate how the introduction of correct examples influences the model's ability to maintain misconception accuracy while recovering overall correct problem-solving skills.

\subsubsection{Recovery of Correct Problem-Solving}

Figure~\ref{fig:weak_csa_a_vs_na_strong_csa} illustrates the remarkable recovery of overall correct problem-solving abilities in our CSMs as we introduced correct training examples. For most misconceptions, we observed a sharp increase in OCA with the introduction of just a small number of correct examples. For instance, with M1, OCA increased from 23.15\% (with no correct examples) to 87.43\% with just 100 correct examples ($n_c/n_m = 0.25$). Similarly, for M11 (incorrectly combining terms), OCA rose from 26.39\% to 88.34\% with 200 correct examples ($n_c/n_m = 0.25$).

Notably, in many cases, a ratio of $n_c/n_m = 0.25$ was sufficient to restore a high level of OCA, suggesting an efficient frontier in balancing misconception and correct example training. As we increased this ratio further, we observed diminishing returns. For example, for M1, the gain in OCA from $n_c/n_m = 0.25$ to $1.0$ was only about 3 percentage points (from 87.43\% to 90.81\%). These results demonstrate our ability to satisfy Property 2 of CSMs while maintaining Property 1.

\subsubsection{Retention of Misconception Accuracy}

Importantly, the introduction of correct examples did not significantly impair the model's ability to retain misconceptions. For most misconception types, we saw an insignificant decrease in Misconception Accuracy compared to the models trained solely on misconception examples. For instance, M1 maintained an MA above 90\% even with $n_c/n_m = 1.0$. M11 showed remarkably stable performance, with MA increasing from 92.97\% to 98.44\% as $n_c/n_m$ increased from 0.25 to 1.0. 

\textbf{These findings demonstrate that with the strategic introduction of correct samples with misconception samples, it is possible to create models that satisfy both properties of CSMs.}

%% file: sections/conclusion.tex
This study makes two significant contributions to the field of AI in education: the development of MalAlgoPy, a novel library for modeling algebraic misconceptions, and the introduction of Cognitive Student Models (CSMs). Using MalAlgoPy, we demonstrated that LLMs can effectively learn to replicate common algebraic misconceptions. However, our initial experiments revealed a critical challenge: as models became proficient at replicating misconceptions, they lost their ability to solve problems correctly in other contexts. We addressed this issue by strategically introducing correct examples alongside misconception training, achieving a balance that allows CSMs to both replicate misconceptions and maintain correct problem-solving abilities. These findings have important implications for the development of more nuanced and effective AI-driven educational technologies.

%% file: sections/limitations.tex
This study, while providing valuable insights, has two main limitations. Firstly, as an initial investigation, our focus was solely on algebraic problem-solving, which limits the generalizability of our findings to other mathematical domains or subjects. Future research should explore the applicability of these models to a broader range of mathematical concepts. Secondly, our current implementation of MalAlgoPy is optimized for single-variable linear equations, which, while covering a significant portion of elementary algebra, does not include more complex algebraic structures such as systems of equations or polynomial factorization. Expanding the library to encompass these areas could provide a more comprehensive understanding of LLMs' capabilities in modeling student cognition across various levels of algebraic complexity.

%% file: sections/appendix.tex

\input{sections/appendix_exp}

%% file: sections/appendix_exp.tex
\subsection{Misconception Acquisition Variability in CSMs}

Our analysis revealed significant variability in how different misconceptions were acquired by the CSMs, as illustrated in Figure \ref{fig:app_misconception_vs_correct_accuracy_4plots}. We also observed distinct patterns in the relationship between Misconception Accuracy (MA) and Overall Correct Accuracy (OCA) as the number of misconception training examples increased:

\paragraph{Rapid Acquisition:} Some misconceptions, such as M1 (distributing multiplication over parentheses incorrectly) and M11 (incorrectly combining terms), showed rapid acquisition. For M1, MA increased from 0.81\% at 100 examples to 89.65\% at 800 examples, while OCA dropped from 53.98\% to 23.15\% over the same range.

\paragraph{Gradual Acquisition:} Other misconceptions like M14\_M15 demonstrated more gradual learning curves. MA for M14\_M15 increased steadily from 1.04\% at 100 examples to 98.05\% at 3200 examples, with OCA decreasing from 63.72\% to 18.83\%.

\paragraph{Resistant Acquisition:} Interestingly, some misconceptions proved resistant to acquisition. M4 and M5 stood out, with M4 reaching only 9.38\% MA at 3200 examples, while its OCA dropped from 79.50\% to 18.77\%.

\paragraph{Immediate High Acquisition:} Certain misconceptions, like M20\_S20, showed immediate high acquisition, achieving 94.52\% MA with just 100 examples, while OCA remained relatively stable, only dropping to 24.48\% at 800 examples.

\paragraph{Trade-off between MA and OCA:} Across most misconceptions, we observed an inverse relationship between MA and OCA. As MA increased with more training examples, OCA generally decreased, though the rate and extent of this trade-off varied among misconceptions.

These findings highlight the complex nature of misconception acquisition in CSMs and underscore the challenges in simultaneously satisfying both properties of CSMs as discussed in Section \ref{sec:csa}. The variability observed emphasizes the need for careful calibration of misconception and correct examples in training, as explored in our subsequent experiments on balancing these properties.

\input{figures/app_misc_acc}
\input{figures/misc_w_correct}
\input{figures/split_acc}

\subsection{Analysis of MA Retention and OCA Recovery Variability across Misconceptions}
\label{app:misc_scm}

Figure \ref{fig:correct_solving_accuracy_two_plots} illustrates the effects of introducing correct examples on both Misconception Accuracy (MA) and Overall Correct Accuracy (OCA) across various misconceptions. This analysis expands on our findings from Section \ref{sec:exps}, providing a comprehensive view of how CSMs maintain both properties simultaneously.

\paragraph{Retention of Misconceptions:} Across most misconceptions, we observed remarkable retention of MA (solid lines) even with the introduction of substantial correct examples. For instance, M1 maintained an MA above 90\% even as the number of correct examples increased to 800. This resilience suggests that once acquired, misconceptions remain stable in the model's knowledge representation.

\paragraph{Recovery of Correct Problem-Solving:} Simultaneously, we observed a significant recovery in OCA (dashed lines) with the introduction of correct examples. For most misconceptions, OCA shows a sharp increase from the initial `S' point, often reaching above 80\% with as few as 100-200 correct examples. This rapid recovery demonstrates the model's ability to relearn correct problem-solving strategies without compromising acquired misconceptions.

\paragraph{Variability Across Misconceptions:} While the general trend of MA retention and OCA recovery is consistent, we observed variability in the specific patterns across different misconceptions. For example, M4 and M5 show distinctly different behaviors, with lower MA and a more gradual increase in OCA compared to other misconceptions.

These findings provide evidence that it is indeed possible to create CSMs that satisfy both properties simultaneously, maintaining high MA while recovering high OCA. The ability to balance these properties through strategic introduction of correct examples offers a promising approach for developing more realistic and nuanced student models.

\subsection{Misconception-wise Variability in Impact on Applicable and Non-Applicable Problem Types}

Figure \ref{fig:avg_accuracy_A_vs_NA_4plots} illustrates the misconception-specific impact of increasing training examples on Correct Accuracy for both Applicable ($CA_A$) and Non-Applicable ($CA_{NA}$) problem types. Starting from initial accuracies around 90\% for both metrics, we observe significant variability in how different misconceptions affect model performance:

\paragraph{Differential Rates of Decline:} All misconceptions show a decline in both $CA_A$ and $CA_{NA}$, but with varying rates. M1 exhibits a precipitous drop in $CA_A$ from 31.1\% to nearly 0\% within 400 examples, while its $CA_{NA}$ decreases more gradually from 64.1\% to 19.8\% over 3200 examples. M6 shows a more moderate decline, with $CA_A$ dropping to around 39.5\% and $CA_{NA}$ to 58.5\% at 3200 examples.

\paragraph{Varying Gaps Between $CA_A$ and $CA_{NA}$:} The difference between $CA_A$ and $CA_{NA}$ evolves differently across misconceptions. For M12\_S15\_M13, the gap widens as training progresses, with $CA_A$ plummeting to near 0\% while $CA_{NA}$ settles around 55\%. In contrast, M14\_M15 shows a narrowing gap, with both metrics converging around 20\% at 3200 examples.

\paragraph{Anomalous Patterns:} Some misconceptions display unique patterns. M17\_M18 maintains relatively high $CA_A$ (above 85\%) for the first 200 examples before rapidly declining, suggesting a threshold effect in misconception acquisition. M4 and M5 show less severe degradation in $CA_A$, retaining values around 3-4\% even at 3200 examples.

\paragraph{Relative Resilience in $CA_{NA}$:} While both $CA_A$ and $CA_{NA}$ decline across all misconceptions, $CA_{NA}$ generally shows less severe degradation. However, this relative resilience varies. M2\_S3 maintains $CA_{NA}$ at 49.7\% at 3200 examples (compared to 4.3\% for $CA_A$), while M17\_M18 sees $CA_{NA}$ drop to 13.4\% (with $CA_A$ at 0.8\%). It's important to note that despite this relative resilience, $CA_{NA}$ still experiences substantial decline from its initial high accuracy.

This misconception-wise analysis reveals the complex and varied ways in which different misconceptions affect model performance. It highlights that while misconception training generally degrades performance on both applicable and non-applicable problem types, the extent and pattern of this degradation vary significantly across misconceptions. This variability underscores the need for nuanced, misconception-specific approaches in developing and evaluating Cognitive Student Models.

%% file: figures/app_misc_acc.tex
\begin{figure*}[t!]
\centering
\includegraphics[width=\linewidth]{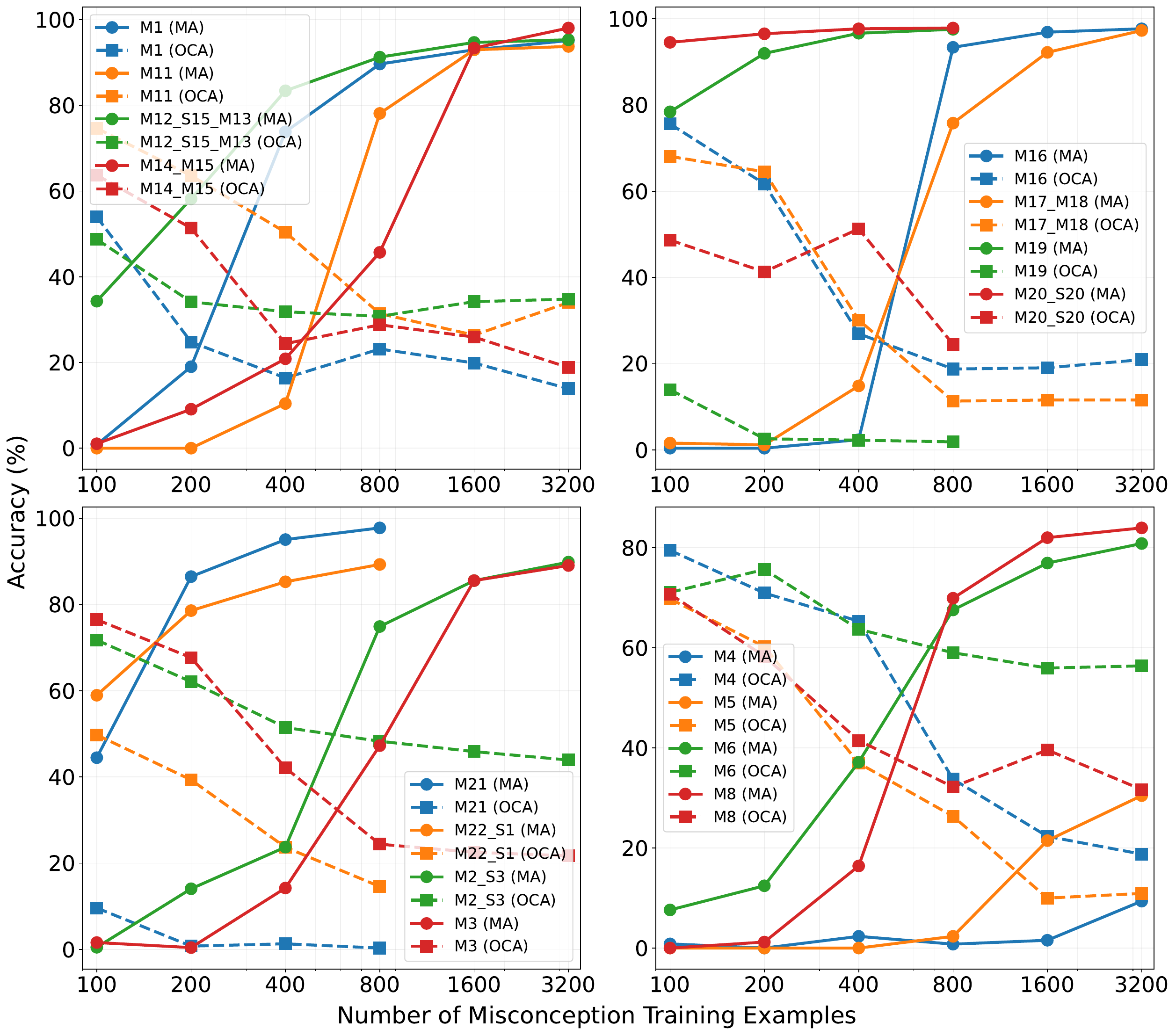} 
\hfill
\caption{Learning dynamics of misconceptions in Cognitive Student Models (CSMs).Solid lines represent Misconception Accuracy (MA), while dashed lines show Overall Correct Accuracy (OCA). As the number of misconception training examples increases, MA generally rises while OCA decreases, illustrating the trade-off between acquiring misconceptions and maintaining correct problem-solving abilities. The four subplots display results for different sets of misconceptions, each represented by a unique color. This visualization demonstrates the variability in acquisition patterns across different misconception types and highlights the challenge in developing CSMs that accurately replicate specific errors while preserving general problem-solving skills.}
\label{fig:app_misconception_vs_correct_accuracy_4plots}
\end{figure*}

%% file: figures/misc_w_correct.tex
\begin{figure*}[t!]
\centering
\includegraphics[width=\linewidth]{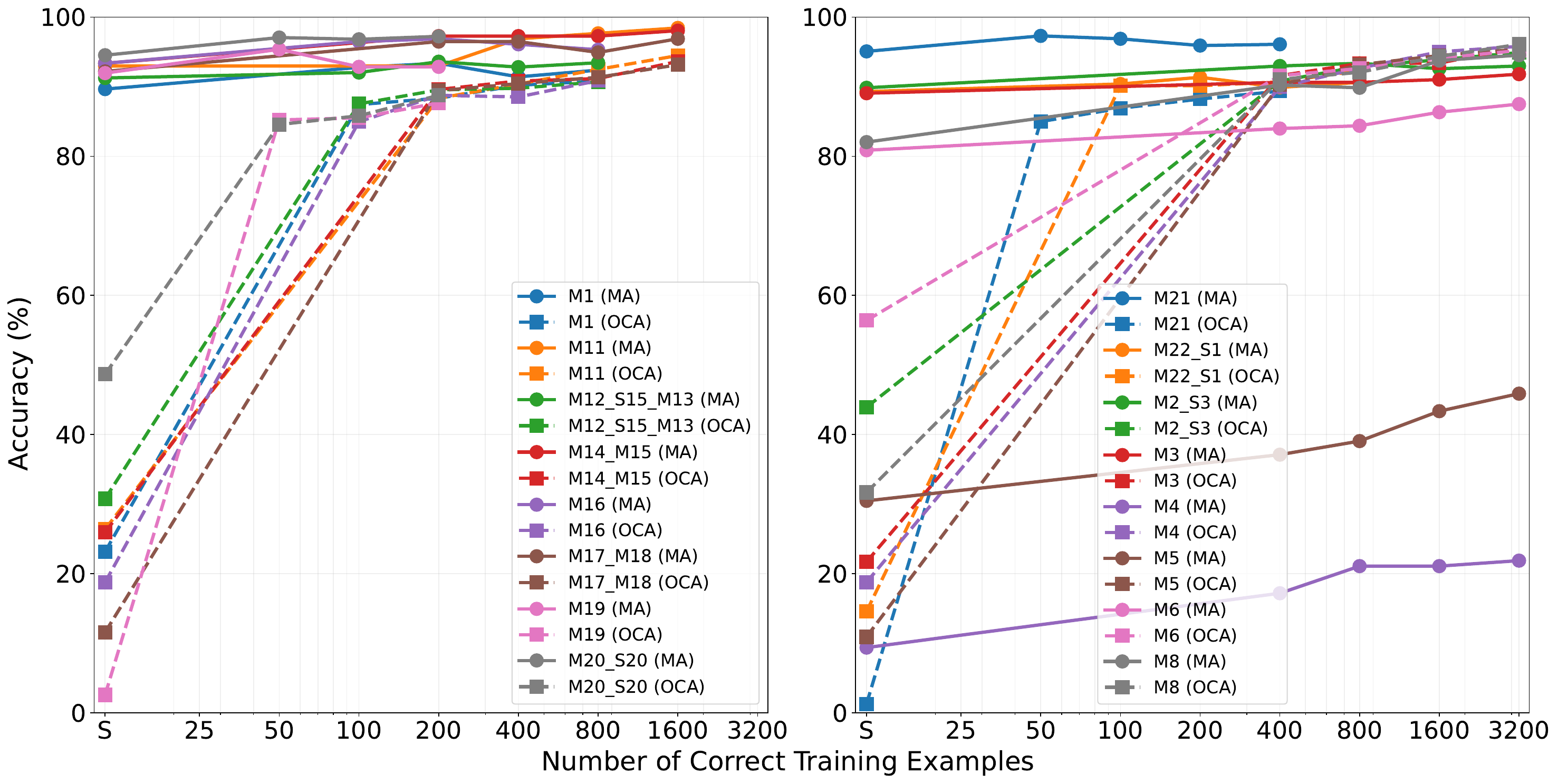} 
\hfill
\caption{Recovery of correct problem-solving abilities with introduction of correct examples. This graph shows how adding correct training examples alongside misconception examples helps the model regain correct solving accuracy. The rapid improvement suggests an effective method for creating balanced student models that can replicate misconceptions while retaining correct problem-solving skills.}
\label{fig:correct_solving_accuracy_two_plots}
\end{figure*}

%% file: figures/split_acc.tex
\begin{figure*}[ht!]
\centering
\includegraphics[width=\linewidth]{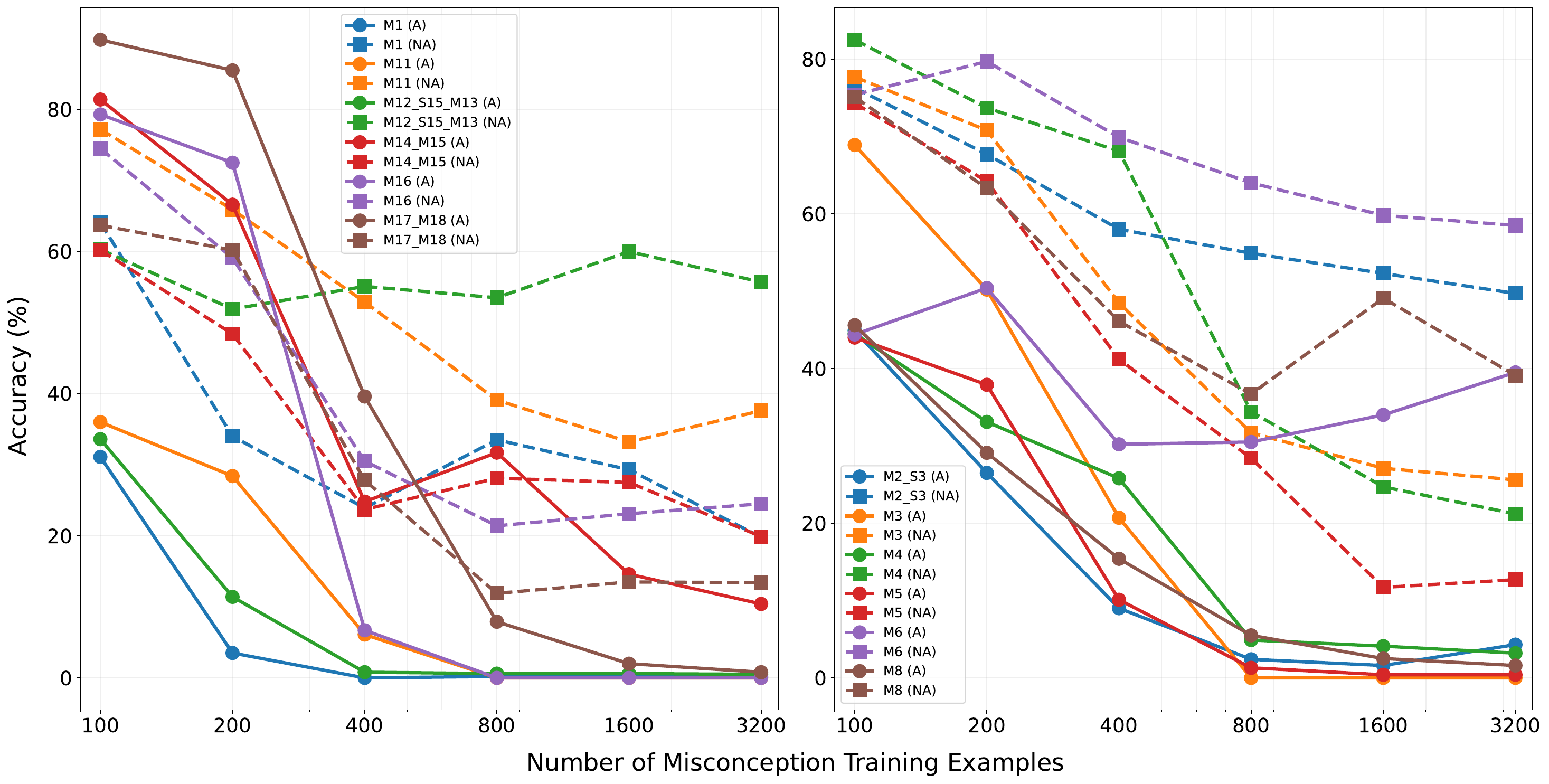} 
\hfill
\caption{Impact of misconception learning on problem types applicable (A) and non-applicable (NA) to the misconception. Solid lines represent accuracy for types where the misconception is applicable, while dashed lines show types where it's not. The goal is to create a model that exhibits learned misconceptions for A problems while maintaining correct solving abilities for NA problems, mimicking real student behavior.}
\label{fig:avg_accuracy_A_vs_NA_4plots}
\end{figure*}